\documentclass{article}

\usepackage{amssymb}
\usepackage{amsfonts}

\title{
The phenomenology of
neutrinos with Majorana mass terms
and standard-model interactions
derived in the charge-parity basis}

\author{R. Plaga \\ Federal Office for Information Security (BSI), D-53175 Bonn, Germany \\ rainer.plaga@bsi.bund.de}

\begin{document}
\maketitle
\begin{abstract}

The physical mechanisms that make a neutrino with standard-model (SM) weak interactions
(``standard-model-interaction
(SMI) neutrino'')
a ``lepton-number conservation (LNC)
violating'' neutrino such as the Majorana neutrino are analysed
in a basis of two Majorana states that have
opposite charge-parity (``charge-parity basis'').
\\
It is necessary to assume that Majorana neutrinos interact with a certain
weak-interaction Hamiltonian ``H$_{PE}$'' to prove that they have
the same phenomenology (to first order) as the
SMI-neutrino in the limit m $\rightarrow$ 0. But H$_{PE}$
violates lepton-number conservation and is therefore
qualitatively different from 
the SM Hamiltonian ``H$_{SM}$''. Because even a Majorana neutrino
is nowadays believed to interact with SM-interactions, H$_{PE}$ is excluded. 
This means that the above necessary assumption for the proof of
the ``Dirac-Majorana confusion theorem'' can no longer be made.
\\
Non standard-model Majorana mass terms modify the equation
of motion of the neutrino by being different in sign and/or
value for the two components of the charge-parity basis. 
A small Majorana mass that is larger than any Dirac mass
makes the neutrino not a Majorana but a ``pseudo-Majorana'' 
particle that has no definite
chirality and therefore has a different phenomenology than the physical neutrino.
A combination of a large Majorana and
Dirac mass of nearly equal value makes the neutrino a Majorana neutrino.
However if this Majorana neutrino has
SM interactions, its weak transition amplitudes are a factor
$\sqrt{2}$ smaller than the ones observed for the physical neutrino.
Only with 
a small Dirac mass that is larger than any Majorana mass (and in the massless case),
the physical neutrino's phenomenology is correctly predicted by the SM.
Such a mass combination makes the neutrino a Dirac- or (the most likely possibility for the physical neutrino) 
Pontecorvo's pseudo-Dirac particle which features 
neutrino-antineutrino oscillations, 
that violate LNC.
Pseudo-Dirac neutrinos 
enable a completely negligible rate for neutrinoless double-beta decay if
there is no Majorana-mass independent
decay mechanism.
\\ 
Off-diagonal components of the mass matrix in the charge-parity basis
make the neutrino a mixture of Dirac field with a different particle
and anti-particle mass (i.e. a mass that violates $\mathcal{CPT}$ invariance)
and a pseudo-Dirac field. Such a neutrino leads to a phenomenology 
similar to the one with additional generations of sterile neutrinos.
\end{abstract}

\section{Introduction}
The question whether the physical neutrino is a Majorana
particle, is one of the major open problems
in particle physics\cite{moha}. 
The standard-model of particle physics\cite{quigg} is not
believed to be a complete theory, but to be correct to very good approximation.
Therefore, while it
is possible that the
neutrino has extra properties that are not contained
in the standard model (like e.g. Majorana mass terms),
it is generally believed that
its weak interaction is described, at least to
good approximation, by the standard model.
Let us call a neutrino with this property
``standard-model-interaction (SMI) neutrino''.
\\
This paper analyses 
the phenomenology of the SMI-neutrino with a
Majorana mass term for the first time systematically in the
``charge-parity basis''. The elements of this
basis are two Majorana states, and therefore it seems
well adapted to study this question.
The aim is to fully understand 
the physical mechanisms that 
can ``make'' the SMI-neutrino
a lepton-number conservation (LNC) violating neutrino such as
the Majorana neutrino. The analysis is
performed only for a single flavor because flavour mixing is
not its topic.
The full second quantised field theory is used throughout, because
a first quantised treatment of Majorana neutrinos is impossible.
\\
In section \ref{defins} I critically
review the definition 
of a ``Majorana neutrino'',
introduce the charge-parity basis
and formulate the definition in this basis.
The physical mechanisms that
can induce a SMI neutrino to violate 
LNC
are analysed in section \ref{mech}. 
In section \ref{offaxis} 
I will work out the phenomenology
of neutrinos with mass terms that couple the components of
opposite charge parity. 
Section \ref{sum} summarises the novel theoretical
results and section \ref{concl} explains what they
mean for the ``physical'', i.e. really existing, neutrino.

\subsection{Preview: which widely held beliefs are put into question
and where these contradictions are resolved}
The results of the paper contradict two widely held
beliefs. 
\\
The first is that a ``Majorana-Dirac confusion theorem'' applies to the physical
neutrino. I will explain why the confusion theorem can only be proved 
under the assumption that Majorana neutrinos have
an exotic, non-SM weak interaction
in section \ref{confuth}. 
This assumption was only tenable until it
was generally accepted that the SM describes the neutrino's
weak interaction to good approximation, and therefore
the confusion theorem must not necessarily hold any more.
This insight clears the way to not reject out of hand
a straightforward demonstration 
that the physical neutrino is no
Majorana neutrino (in section \ref{largm}), ``because it contradicts
the confusion theorem''.
\\
The second belief is that 
a small Majorana mass makes neutrinos Majorana particles that
are not their own charge conjugate and that in the 
basis of Majorana states the mass matrix has
Majorana masses on the diagonal and Dirac masses
as the off-diagonal elements. I will
demonstrate in section \ref{pM}:
\\
- this
belief is absolutely correct, however
not for the Majorana neutrino but rather
for states that I
christen ``pseudo-Majorana'' and the basis formed with them
\\
- the physical
neutrino is not a pseudo-Majorana neutrino.
\\
In the charge-parity basis both Dirac and Majorana
masses remain on the diagonal of the mass matrix 
and its off-diagonal elements
violate $\mathcal{CPT}$ invariance (section \ref{offaxis}).

\section{The definition of Majorana neutrino fields and the charge-parity basis}
\label{defins}
A Dirac neutrino field $\nu_D$ at a position $x,t$ in space-time can be written in compact form as\cite{weinberg1,bd}:
\begin{equation}
\nu_D(x,t) = {\sum\limits_{k}}^{(+)} b_k^{\dagger} v_k + {\sum\limits_{k}}^{(-)} d_k u_k
\label{modes}
\end{equation}
Here the sum extends over all momentum and spin states (``modes'') 
the field can be in.
The (+) symbolises the modes with positive energy and (-) the modes
with negative energy, i.e. modes that are reversed in time in the
sense that instead of a particle or antiparticle ``moving into'' mode k (being created, symbolised by 
the +), a
particle or antiparticle
is ``removed from'' mode k (being annihilated, symbolised by the -).
$b_k^{\dagger}$ is the creation operator for a particle (symbolised by the $b$) with spinor $v_k$ in mode k,
and  $d_k$ is the annihilation operator for an antiparticle (symbolised by the $d$) with spinor $u_k$ in mode k.
Below ``$\nu_x$'' symbolises a neutrino field in a state $x$ and $|\rangle$ a state vector that
contains the amplitudes of the states a system can be in.
$|\nu_x\rangle$ is a shorthand for a system that has amplitude 1 for being in the neutrino
state $\nu_x$.
\\
In the Majorana representation of the $\gamma$ matrices\cite{itz},
which I will use throughout this paper (see appendix \ref{rep} for
explicit $\gamma$ matrices I chose), the 
operation of ``charge conjugation'' is defined as
taking the Hermitian conjugate of the field\cite{bd}:
\begin{equation}
{{\nu}}(x,t)^c = {{\nu}}(x,t)^{\dagger T}  \ \ \ ({\rm Majorana \ representation})
\label{def2}
\end{equation}
The transposition operator T is
to be applied only to the spinors ($u,v$) , but not to the creation and annihilation operators
\cite{sakurait}. Therefore charge conjugation turns the spinor column $u$
to a spinor column $v$, and e.g. a creation operator $b^{\dagger}$ into
an annihilation operator $b$\cite{bd}.
Applying eq.(\ref{def2}) to eq.(\ref{modes}) one obtains:
\begin{equation}
\nu_D^c(x,t) = {{\sum\limits_{k}}^{(-)} b_k u_k + {\sum\limits_{k}}^{(+)} d_k^{\dagger} v_k}
\label{modesc}
\end{equation}
Comparing eq.(\ref{modes}) and eq.(\ref{modesc}) 
one finds that charge conjugation is fully
characterised by: ``replace $b$ with $d$ and $b^\dagger$ with $d^\dagger$''.
For the correct description of charged fermions, like electrons,
it is necessary to make the fundamental 
assumption that the operators $b^{\dagger}$ and 
$d^{\dagger}$ are qualitatively different. In other words,
one has to assume that there are two types
of fundamental fields ``particles'' and ``antiparticles''.
Majorana asked\cite{maj}:
might for neutrinos $b^{\dagger} \equiv d^{\dagger}$?
A neutrino with this property, a ``Majorana neutrino'' $\nu_M$, is self-charge conjugate
i.e. it fulfils the ``Majorana condition''\cite{bilenky}:
\begin{equation}
{\nu_M}(x,t)^c = e^{i\alpha} {\nu_M}(x,t)
\label{def}
\end{equation}
Here $e^{i\alpha}$ is a phase factor which is called ``{\it the charge-parity of the field}''.
\\
There are two widespread misconceptions about this definition.
\\
The first misconception is that charge conjugation flips chirality, i.e. a
left-chiral neutrino becomes a right-chiral antineutrino under charge
conjugation\footnote{The author was under the spell of this misconception
himself. I am very much indebted
to P. Pal for explaining me
this point in detail (see also Ref.\cite{palash}).}. 
It is in principle well known\footnote{See e.g. fig.3.5
in the classic textbook of Perkins\cite{perkins}, that shows
that the neutrino helicity does not change under charge conjugation. But
in the massless limit helicity and chirality are identical (see e.g.\cite{palash} for a
proof), so that neither changes chirality under charge conjugation.} that
charge conjugation has no effect whatsoever on the
spatial and spin modes k. 
In particular a field creating a (massless) neutrino
with chirality=-1 (left-chiral neutrino) is transformed
by charge conjugation into a field creating an antineutrino with 
unchanged chirality = -1.
There is confusion about this fact in the literature, presumably
because a spinor ${(1-\gamma_5) \over 2} v(x)$ = $v_L$ is indeed turned into a spinor state
${(1 + \gamma_5) \over 2} u(x)$ = $u_R$ with opposite chirality by charge conjugation\footnote{For a spinor in the Majorana representation chosen here, charge conjugation is complex conjugation. $\gamma_5$ is purely imaginary in the Majorana representation. Therefore charge conjugation turns 1-$\gamma_5$ to 1+$\gamma_5$.}: $v_L^c$ = $u_R$\cite{bd}. 
However, if the left-chiral neutrino {\it field} state:
\begin{equation}
{\nu_{DL}}(x,t) = {\sum\limits_{k}}^{(+)} b_k^{\dagger} v_{kL} + {\sum\limits_{k}}^{(-)} d_k u_{kR}
\end{equation}
is charge conjugated, {\it both} the spinor and the creation operator
and annihilation are charge conjugated and one obtains:
\begin{equation}
{\nu_{DL}}(x)^c = {\sum\limits_{k}}^{(-)} b_k u_{kR}(x) + {\sum\limits_{k}}^{(+)} d_k^{\dagger} v_{kL}(x)
\end{equation}
i.e. indeed
a left-chiral neutrino field is charge-conjugated to a left-chiral antineutrino
field. Charge conjugation {\it only} affects the ``particle-antiparticle'' character
of the field.
\\
The second misconception is a
claim \cite{kayser82,kayser} that ``dressed'' (a fancy expression for ``weakly interacting'')
neutrinos cannot be self-charge conjugate
i.e. cannot fulfil eq.(\ref{def}) ``because the weak interaction
is not invariant under charge conjugation''. 
The weak interaction of a left-chiral Majorana neutrino $\nu_{ML}$ is not
invariant under charge-conjugation because
e.g. the virtual reaction $\nu_{ML} \rightarrow$ e$^-$ W$^+$ 
takes place but its charge conjugate 
$\nu_{ML} \rightarrow$ e$^+$ W$^-$ does not.
The fact that the field $\nu_{ML}$ does not
change under charge conjugation does not invalidate
the fact that its weak reactions are not charge-conjugation
invariant. Therefore the non-invariance of the reaction under charge-conjugation
is not in contradiction to a self-charge conjugacy of the neutrino.
See appendix
\ref{app1} for a detailed refutation of a published 
proof for the above claim and further comments.
\\
I now introduce the {\it charge-parity basis} consisting of two Majorana states
$\nu_+$ and $\nu_-$.
Below  
I will often drop the arguments $x,t$
and often also a subindex R,L indicating the
chirality of the neutrino state, because
the set of states of greatest interest for the present analysis has
just the elements of $\nu$ (eq.(\ref{modes})) 
and $\nu^c$ (eq.(\ref{modesc})).
$\nu_+$ is defined as:
\begin{equation}
{{\nu}_+} = {1 \over \sqrt{2}}(\nu_D + \nu_D^c)
\label{cmaje+}
\end{equation}
which fulfils eq.(\ref{def}) with a phase (or charge-parity) e$^{i\alpha}$= +1:
\begin{equation}
{{\nu}_+}^c = \nu_+.
\label{majc1}
\end{equation}
The other component is:
\begin{equation}
{{\nu}_-} = {1 \over \sqrt{2}}(\nu_D - \nu_D^c).
\label{cmaje-}
\end{equation}
which possesses a charge-parity of -1:
\begin{equation}
{{\nu}_-}^c = - \nu_-
\label{majc2}
\end{equation}
Obviously $\nu_+$ and $\nu_-$ are both Majorana states $\nu_M$ because
they fulfil eq.(\ref{def}).
We can then describe the Dirac neutrino as
\begin{equation}
{\nu_D} = {1 \over \sqrt{2}}(\nu_+ + \nu_-),
\label{dirac1}
\end{equation}
and the Dirac antineutrino as
\begin{equation}
{{\nu}_D}^c = {1 \over \sqrt{2}}(\nu_+ + \nu_-)^c= {1 \over \sqrt{2}}(\nu_+ - \nu_-) \neq {\nu}_D.
\label{cdirac}
\end{equation}
In order to formulate the definition of the Majorana neutrino in the charge-parity
basis we need to permanently restrict the set of states available to the neutrino
to a subset of fields that fulfils the Majorana condition eq.(\ref{def}).
The Majorana condition
eq.(\ref{def}), i.e. definition of the Majorana neutrino evidently always holds if the
\begin{itemize}
\item
{\bf Majorana restriction :}
\\
{\it The neutrino state e$^{i\alpha}$ $\nu_-$ (or equivalently e$^{i\alpha}$ $\nu_+$) is permanently excluded from the physically accessible Hilbert space.}
\end{itemize}
is guaranteed by some physical mechanism (for one angle $\alpha$).
If the Majorana restriction holds in general, the neutrino ``is'' a Majorana field.

\section{Mechanisms that ``make'' the neutrino a Majorana neutrino,
or another LNC violating neutrino}
\label{mech}
Originally it was simply assumed that the neutrino
might ``be'' a Majorana neutrino in the sense that the statement
``the state $\nu_-$ does not exist'' (i.e. the Majorana restriction)
is a law of nature. However, as we discuss in detail below, in the
unmodified standard model the neutrino is a Dirac neutrino,
and according to eq.(\ref{dirac1}) then both states $\nu_+$
and $\nu_-$ do exist.
Therefore, to make the neutrino a Majorana particle, 
some mechanism beyond the SM needs to implement the Majorana restriction.
Which mechanisms can exclude components of the charge parity basis from the
accessible Hilbert space without violating the above
assumption that the SM describes the weak interaction to
good approximation?
In principle there are two possibilities:
\\
1. Non-SM weak interactions 
of the field
\\
2. Non-SM masses of the field
\\
I will now analyse both possibilities in turn.
\subsection{Possibility 1:Weak interaction Hamiltonians that make Majorana neutrinos}
\label{winter}
\subsubsection{SMI-neutrinos cannot be made Majorana neutrinos by the weak interaction}
\label{smi3}
Can the weak interaction create a massless (or only Dirac-massive) 
Majorana neutrino?
As discussed in the Introduction,
today there is a broad consensus that the neutrino's charge
current Hamiltonian 
is the one described by the standard model i.e.
\begin{equation}
H_{SM} = {g \over 2\sqrt{2}} (W^- (\bar{e} \gamma_{\mu} (1-\gamma_5)\nu_D) +
W^+ (\bar{{\nu}}_{D} \gamma_{\mu} (1+\gamma_5) e)).
\label{intsm}
\end{equation}
Clearly this interaction creates a massless (or only
Dirac massive) neutrino 
as $\nu_D$, i.e. as a Dirac particle.
But if a different charged current Hamiltonian, let us call it ``H$_{PE}$'', 
would create e.g. only the field state $\nu_+$, then the field state
$\nu_-$ would be permanently excluded (i.e the Majorana restriction would be fulfilled
for $\alpha$=0) and
a massless neutrino would be created as a Majorana neutrino.
The charged-current interaction term that achieves this is:
\begin{equation}
H_{PE} = {g \over 2\sqrt{2}} (W^- (\bar{e} \gamma_{\mu} (1-\gamma_5)\nu_{+}) +
 W^+ (\bar{\nu}_{+} \gamma_{\mu} (1 + \gamma_5) e).
\label{lmaj}
\end{equation}
It was first discussed by Pauli\cite{pauli57} (his eq.(22)) and Enz\cite{enz}. In eq.(\ref{lmaj})
I replaced a hadronic current in their expression for H$_{PE}$ by the W-boson for simplicity.
I propose to call H$_{PE}$ ``Pauli-Enz''(PE) interaction.
\\
Using eq.(\ref{cmaje+}) $H_{PE}$ can be rewritten as:
\begin{equation}
H_{PE} = {1 \over \sqrt{2}} H_{SM} + {g \over 4} (W^- (\bar{e} \gamma_{\mu} (1-\gamma_5)\nu_D^c) +
W^+ (\bar{{\nu^c}}_{D} \gamma_{\mu} (1+\gamma_5) e)).
\label{lmaj2}
\end{equation}
The SM 
necessarily does
conserve lepton number\cite{kraus} by mapping a positron in a neutrino
and an electron into an antineutrino with a SU(2) transformation.
We see that H$_{PE}$ does not merely add a non-SM 
LNC violating 
second term but that it changes the first SM term by a factor 
${1 \over \sqrt{2}}$. Moreover the LNC violating second term is not
small, but of equal magnitude as the first term.
Therefore, if neutrinos would interact with H$_{PE}$, the SM would
be wrong. In appendix \ref{lncd} I explicitely demonstrate LNC violation in
a weak reaction with H$_{PE}$.
Summarising I formulate the following 
\begin{itemize}
 \item 
{\bf Fact}
\\
{\it If the weak interaction would implement the Majorana
restriction (``make the neutrino Majorana'') - in the sense
that it would produce even a massless neutrino as Majorana
particle - the SM would not describe weak interactions
to good approximation.}
\end{itemize}
Therefore - if the physical neutrino is a Majorana
particle - it must be because of the action of non-SM Majorana mass term (see section \ref{massmake} where we will discuss how
this mechanism keeps the SM intact).
\\
Let us examine the argument of this subsection again, from a 
somewhat different angle. 
Some SM bosons
happen to be ``a priori'' self-charge conjugate, namely the photon,
the Z$_0$ and the Higgs boson (even though they
are not called Majorana fields, because this
designation is reserved for fermions).
The neutrino fields happens not to be self-charge 
conjugate in the SM,
i.e. the SM weak interaction  
produces only Dirac neutrinos and antineutrinos if they
are massless or only Dirac-massive. 
One can consider a theory of weak interactions
with a charged-current Hamiltonian ``H$_{PE}$''
in which massless
neutrinos are produced as self-charge conjugate 
fields (i.e. are self-charge conjugate ``a priori'')
but such a theory cannot be the SM
in which massless neutrinos are Dirac particles because they are obtained by a SU(2)
transformation from electrons, which are necessarily
Dirac particles because they are electrically charged.

\subsubsection{SM and LNC violating weak interactions
of Majorana neutrinos: the confusion theorem}
\label{confuth}
In the late 1950s, long before the SM was established, 
the phenomenology of parity-violating Majorana neutrinos
was clarified in a flurry of papers\cite{pauli57,enz,flurry,radi}.
At that time the possibility
that the weak interaction charged current Hamiltonian is the LNV
violating H$_{PE}$ 
was entirely reasonable. The title
of Enz' Ref.\cite{enz} is: {\it Fermi Interaction with Non-Conservation of ``Lepton Charge'' and of Parity}.
The standard case
of a weak-interaction Hamiltonian
H$_{SM}$ was equally reasonable. (At the time, the expression for H$_{PE}$
and H$_{SM}$ contained a hadronic current instead
of the still unknown W-boson, I will use the modern notation below.)
\\
In the previous subsection \ref{smi3} I excluded the possibility
that H$_{PE}$ is the Hamiltonian of the physical neutrino.
In this subsection I review the consequences of not excluding
it, in order to
understand the origin of the ``confusion theorem''
and why one of the assumptions needed for its proof can no longer
be made.
\\
In the 1950s there was no way to decide a priori if 
H$_{PE}$ and H$_{SM}$ is the correct Hamiltonian and therefore
all authors\cite{flurry,radi} {\it implicitly}
made the
\begin{itemize}
\item
{\bf Assumption PE}
\\
{\it 1.If neutrinos are Majorana particles they
interact with H$_{PE}$ and
\\
2. if they are Dirac neutrinos they interact
interact with a Hamiltonian H$_{SM}$.}
\end{itemize}
The following theorem can be proved if, and only if, assumption PE is true.
\begin{itemize}
\item
{\bf Confusion theorem}
\\
{\it Dirac and Majorana neutrinos have the same phenomenology
in the limit m $\rightarrow$ 0.}
\end{itemize}
This statement was originally called ``equivalence theorem''
by Radicati and Touschek\cite{radi}
and reappeared as ``confusion theorem'' in the more modern literature\cite{kayser}.
Here is its proof:
\\
\
If the neutrino is massless, it is either perfectly right-
or left-chiral. From eqs.(\ref{lmaj},\ref{intsm})
the transition elements 
from a positron and $W^+$ to a massless right-chiral Majorana neutrino
interacting via H$_{PE}$
and to a massless Dirac neutrino interacting via H$_{SM}$ 
(the reactions eq.(\ref{Dreac},\ref{Mreac}) discussed in 
appendix \ref{lncd}) is equal to:
\begin{equation}
{\langle (e^+ W^-)_k |H_{SM}| \bar{\nu}_{DR} \rangle } = 
{\langle (e^+ W^-)_k |H_{PE}| \bar{\nu}_{+R} \rangle } = c \neq 0
\label{tran1r}
\end{equation}
Here the index $k$ symbolises a well defined spatial and spin state of $W^- e^+$.
The transition from an electron and W$^+$ to a right-chiral neutrino is 0 for vanishing mass in both cases:
\begin{equation}
{\langle (e^- W^+)_k |H_{SM}| \bar{\nu}_{DR} \rangle } = 
{\langle (e^- W^+)_k |H_{PE}| \bar{\nu}_{+R} \rangle } = 0
\label{tran1r2}
\end{equation}
The equality of these generic transition amplitudes of a Majorana neutrino 
interacting with H$_{PE}$ and a Dirac neutrino interacting with H$_{SM}$
(and an analogous argument for the weak neutral currents)
prove that Dirac and Majorana neutrino have the same weak interaction
phenomenology {\it if, and only if, assumption PE is true.}
\\
${\langle (e^- W^+)_k |H_{PE}| \bar{\nu}_{+R} \rangle }$ does not vanish
for massive Majorana neutrinos, because massive neutrinos cannot be perfectly
left-chiral.
But ${\langle (e^- W^+)_k |H_{SM}| \bar{\nu}_{DR} \rangle}$
does vanish
even for massive Dirac neutrinos due to lepton-number conservation.
Therefore the confusion theorem only holds in the massless limit. End of proof.
\\
To the best of my knowledge
it was never pointed out {\it explicitely}
that the confusion theorem is only valid if assumption PE is.
In the 1950s no such mention was necessary, because, at the time, PE might well have been
correct and under this circumstance massless Dirac and Majorana neutrinos
could not be discriminated. 
However, nowadays it is believed that
even a Majorana neutrino would have SM weak interactions to good
approximation, i.e. that PE.1 can no longer be made.
Summarising I formulate the following
\begin{itemize}
\item {\bf Fact}
\\
{\it 
If neutrinos are assumed to interact with SM interactions to good approximation, 
assumption PE.1, that is necessary for the proof of the confusion theorem, can no longer be made.}
\end{itemize}
This does not mean that in principle Dirac and Majorana neutrino cannot
have the same phenomenology in the massless limit.
Rather it only means that one cannot conclude from the
confusion theorem that they
have.
In section \ref{largm} I will discuss how the SMI-neutrino can be made a 
light Majorana neutrino
by assigning certain values for the Majorana and Dirac masses.
Based on the above Fact, it will not be possible to simply
conclude that such a Majorana neutrino has the 
weak-interaction phenomenology of the Dirac neutrino
in the limit m $\rightarrow$ 0 ``because of the
confusion theorem''. Rather we
will have to calculate its weak-interaction transition amplitude
to a W$^-$ and e$^+$ and
to compare it with the transition amplitude in  
eq.(\ref{tran1r}).

\subsection{Possibility 2: Mass terms that make pseudo-Majorana, 
pseudo-Dirac and Majorana neutrinos}
\label{massmake}
How can a Majorana mass term make the neutrino
self-charge conjugate, i.e. a Majorana neutrino?
In other words, how can a mass restrict the set of possible field states, i.e. 
implement the Majorana restriction of section \ref{defins}?
In order to answer this question, I will first
derive the equation of motion of a free massive neutrino
in the charge-parity base.
Let us write eq.(\ref{dirac1}) in
vectorial form:
\begin{equation}
\nu_D = {1 \over \sqrt{2}} \left( \begin{array}{c} \nu_+ \\ \nu_-\end{array} \right)
\label{vecN}
\end{equation}
so that the upper component has charge parity +1 and
the lower charge parity = --1.
We now search for the mass matrix in this basis.
\\
I consider
a slightly extended standard-model in which neutrinos can possess ``Dirac 
masses''. These masses are assumed to arise in exactly the same
way as the masses of the charged leptons via coupling
to the Higgs field. The introduction of such masses
has been aptly called ``correcting an oversight'' of the original
SM formulation\cite{quigg}, because there is no known reason
that the coupling of the neutrino to the SM Higgs field must
vanish exactly.
The Dirac mass term in the neutrino Lagrangian is then:
\begin{equation}
L_{m}^{Dirac} =  m_D (\bar{\nu}_D {\nu_D} + \nu_D \bar{\nu}_D)
\label{Diracm}
\end{equation}
Inserting eqs.(\ref{dirac1},\ref{cdirac}) into the mass term eq.(\ref{Diracm}) one gets
in the charge-parity basis:
\begin{equation}
L_m^{Dirac} = m_D (\bar{\nu}_+ \nu_+ + \bar{\nu}_+ \nu_- + \bar{\nu}_- \nu_+ + \bar{\nu}_- \nu_-) 
\label{diracm}
\end{equation}
From the eqs.(\ref{majc1},\ref{majc2}) 
$\bar{\nu}_+ = \nu_+ \gamma_0$ and $\bar{\nu}_- = - \nu_- \gamma_0$
so that the second and third term cancel each other and we are left
with:
\begin{equation}
L_m^{Dirac} = m_D (\bar{\nu}_+ \nu_+ + \bar{\nu}_- \nu_-) = 
(\bar{\nu}_+ \bar{\nu}_-) M_D \left( \begin{array}{c} \nu_+ \\ \nu_-\end{array} \right)
\label{dirmaj}
\end{equation}
with 
\begin{equation}
M_D = \left( \begin{array}{cc}
m_D &   0\\   0 & m_D \\
\end{array} \right).
\label{mass}
\end{equation}
The Dirac mass term of the SMI neutrino is seen to be mathematically equivalent to
mass terms of the two component Majorana fields which have the same
sign and value. 
\\
There are plausible theoretical reasons to suspect that 
non-normalisable
non standard-model Majorana mass terms of the neutrino exist\cite{weinberg}. In the
particle-antiparticle basis they are:
\begin{equation}
L_m^{Majorana} = m_M (\bar{\nu}_D {\nu^c}_D + \bar{\nu^c}_D {\nu_D})
\label{majform2}
\end{equation}
with m$_M$ $\approx$ 10$^{-5}$ - 10$^{-1}$ eV.
Inserting eqs.(\ref{dirac1},\ref{cdirac}) into
eq.(\ref{majform2}) yields in the charge parity basis:
\begin{equation}
L_m^{Majorana} = m_M (\bar{\nu}_+ \nu_+ - \bar{\nu}_- \nu_-) = 
(\bar{\nu}_+ \bar{\nu}_-) M_M \left( \begin{array}{c} \nu_+ \\ \nu_-\end{array} \right)
\label{dirmaj3}
\end{equation}
with 
\begin{equation}
M_M = \left( \begin{array}{cc}
m_M &   0\\   0 & -m_M \\
\end{array} \right).
\label{mmass}
\end{equation}
From this equation and eq.(\ref{dirmaj})
the total mass matrix with both Dirac and
Majorana mass terms is:
\begin{equation}
M_t = M_D + M_M = 
\left( \begin{array}{cc}
m_+ &   0\\   0 & m_- \\
\end{array} \right)
=
\left( \begin{array}{cc}
m_D + m_M &   0\\   0 & m_D - m_M \\
\end{array} \right)
\label{mass2}
\end{equation}
From the Lagrangian of the free neutrino field
\begin{equation}
L = \bar{\nu}_D i \gamma^{\mu} {\partial \over {\partial x^\mu}} \nu_D
+ \bar{\nu}_D M_t {\nu_D}
\label{majlagr2}  
\end{equation}
the equations of motion for $\nu_+$ and $\nu_-$ are:
\begin{equation}
i \gamma^{\mu} {\partial \over {\partial x^\mu}} \nu_+(x,t)
+ (m_D + m_M) \nu_+(x,t) = 0 \ \ \ 
\label{majopl}
\end{equation}
and 
\begin{equation}
i \gamma^{\mu} {\partial \over {\partial x^\mu}} \nu_-(x,t)
+  (m_D - m_M) \nu_-(x,t) = 0 \ \ \
\label{majomi}
\end{equation}
It is now clear in principle how finite Majorana masses can restrict (or enlarge) the state
space of the neutrino: by making the mass term of the
$\nu_+$ and $\nu_-$ component different in value or even sign.
This requires no modification in the weak interactions and together
with the form of eq.(\ref{mass2}) this means that:
\begin{itemize}
 \item 
{\bf Fact}
\\
{\it The Majorana mass terms are a mere non-SM addition to the SM Lagrangian.
They leave the weak-interaction and Higgs-mass part of the
SM completely intact.}
\end{itemize}
To study the effect of this on the neutrino's state space
we have to discriminate between the following
cases:
\begin{enumerate}
\item
m$_D$ and m$_M$ are ``small'' (with respect to the 
smallest energies in laboratory
neutrino reactions, i.e. m$_D$,M$_M$ $\ll$ keV) and
\\
(1.1) m$_M$ $>$ m$_D$ (section \ref{pM}) (opposite sign mass term in eqs.(\ref{majopl},\ref{majomi}))
\\
(1.2) m$_M$ $\leq$ m$_D$ (section \ref{dirstate}) (like sign mass term in eqs.(\ref{majopl},\ref{majomi}))
\item
m$_D$ and m$_M$ are ``large'' with respect
to the largest energies occurring in laboratory
neutrino reactions i.e. m$_D$,m$_M$ $\gg$ 100 GeV but
equal to each other to within a difference smaller than 
the masses of the physical neutrino (section \ref{largm}).
\end{enumerate}
Case 1.1 includes the simplest LNC violating case of a single
small Majorana mass term without a Dirac mass term and
will therefore be discussed first. Contrary to
expectation, it does not make the neutrino a Majorana neutrino.
Case 1.2 is well known to correspond to
a pseudo-Dirac neutrino if m$_D$ $\gg$ m$_M$\cite{giunti},
we will find that this is true also
for m$_D$ $\geq$ m$_M$.
Only Case 2. enforces the Majorana restriction
of section \ref{defins} and makes the SMI neutrino a Majorana particle.
\subsubsection{Small m$_D$ $<$ m$_M$: pseudo-Majorana neutrinos}
\label{pM}
If m$_M$ is ``small'' (in the above sense) and larger than m$_D$
the mass terms in eq.(\ref{majopl}) and eq.(\ref{majomi}) have an
opposite sign.
It is well known that the sign of the mass term in
the Dirac equation is
reversed by multiplying the field by $\gamma_5$,
because $\gamma^{\mu}$ and $\gamma_5$ anticommute\cite{sakurai}.
If the sign of the mass terms in eq.(\ref{majopl}) 
and eq.(\ref{majomi}) the same 
(as it is for case 1.2), it has no effect on observable physics,
because $\gamma_5$ is then just an overall phase factor of both
$\nu_+$ and $\nu_-$. However, if the sign is different, as assumed
for this subsection, $\nu_+$ and $\nu_-$ obtain a {\it relative}
phase of $\gamma_5$ which has observable effects.
Independent of the special form of the spatial state (which is of no concern for our discussion)
the joint solutions of the system of equations eq.(\ref{majopl}) and eq.(\ref{majomi}) can have
either have the following ``particle''-solution form:
\begin{equation}
{\nu}_{pM+} = {1 \over \sqrt{2}}({\gamma_5 \nu_+} +  {\nu_-}) = {1 \over 2}((1 + \gamma_5) \nu - (1 - \gamma_5) \nu^c)
\label{pM+}
\end{equation}
or the corresponding ``antiparticle''-solution form:
\begin{equation}
{\nu}_{pM-} = {1 \over \sqrt{2}}(\gamma_5 \nu_+ -  {\nu_-}) = {1 \over 2}((1 + \gamma_5) \nu^c - (1 - \gamma_5) \nu)
\label{pM-}
\end{equation}
The right-hand form of the equation is obtained by inserting eqs.(\ref{cmaje+},\ref{cmaje-}).
Taking into account that charge conjugation does not flip chirality (see discussion in section \ref{defins})
one immediately concludes from the right-hand side that
charge conjugation transforms ${\nu}_{pM+}$ to
${\nu}_{pM-}$. Therefore ${\nu}_{pM}$ does {\it not} fulfil eq.(\ref{def})
and is no Majorana field, in spite of
containing neutrino and antineutrino components of equal size.
The reason is that the restriction is to a field where a phase factor
$\gamma_5$ appears where a phase factor i would have been necessary
to effect the Majorana restriction of section \ref{defins} with
$\alpha$=$\pi$\footnote{$\gamma_5$ ``comes close'', it is purely
imaginary in the Majorana representation.}.
The limit m$_M \rightarrow$ 0 is not smooth, because
the phase shift of $\gamma_5$ appears for arbitrarily small
m$_M$ but not for m$_M$=0.
\\
Because ${\nu}_{pM+}$ contains both neutrino and antineutrino components,
just like a Majorana neutrino does (eq.(\ref{dirac1})),  
I christen it ``pseudo-Majorana'' neutrino.
${\nu}_{pM+}$ is composed of the components that can both interact weakly
according to H$_{SM}$, i.e. it is the ``active component'', whereas
${\nu}_{pM-}$ is sterile.
From eqs.(\ref{pM+},\ref{pM-})
the pseudo-Majorana neutrino
has no definite chirality
because it has components
of left- and right chirality (remember from
section \ref{defins} that the chirality of a field does
not depend on whether it is particle or antiparticle). 
In the massless limit
chirality = helicity, i.e. the pseudo-Majorana neutrino has
no definite helicity in the massless limit. But the physical
neutrino was of course experimentally determined to have
a helicity = 1 in a state that
is massless to good approximation under the sole assumption that angular momentum
is conserved\cite{goldhaber}. Summarising:
\begin{itemize}
\item
{\bf Fact}
\\
{\it The physical neutrino is
definitely no pseudo-Majorana neutrino (that has no definite chirality),
because it is experimentally
observed to have helicity = 1 in the massless limit 
and therefore has a chirality = 1 to good approximation.} 
\end{itemize}
The pseudo-Majorana neutrino possesses two other properties
that were occasionally erroneously assigned to Majorana neutrinos.
Below I will explain them.
\\
Firstly pseudo-Majorana neutrinos are self-$\mathcal{CPT}$
conjugate in the following sense:
the $\mathcal{CPT}$ transforms\cite{bd} of ${\nu}_{pM+}$:
\begin{equation}
{\nu}_{pM+}^{'} = \mathcal{CPT} {\nu}_{pM+}(\mathcal{CPT})^{-1} = {i \over \sqrt{2}}({\nu_+} -  \gamma_5 {\nu_-}) 
\label{cptpM+}
\end{equation}
would also interact weakly with SM interactions.
Therefore a counterfactual, ``physical'' pseudo-Majorana
neutrino would be in state of incoherent mixture with
the density matrix:
\begin{equation}
\rho(\nu_{pM}) = {1 \over \sqrt{2}} \left(|{\nu}_{pM+}\rangle\langle{\nu}_{pM+}| + |{\nu}_{pM+}^{'}\rangle\langle{\nu}_{pM+}^{'}|\right).
\end{equation}
which is $\mathcal{CPT}$ self-conjugate. The $\mathcal{CPT}$ self-conjugacy 
is seen to be the result of the simple fact that it makes no 
difference whether the phase shift $\gamma_5$ is applied to $\nu_+$
or $\nu_-$.
\\
Secondly in the ``active-sterile'' basis with the components $\nu_{pM+}$ and $\nu_{pM-}$
\begin{equation}
\nu_{pM} = {1\over\sqrt{2}} \left( \begin{array}{c} \nu_{pM-} \\ \nu_{pM+} \end{array} \right).
\label{vecN2}
\end{equation}
(but not the charge parity basis)
the neutrino has a mass matrix with m$_M$ on the diagonal
and m$_D$ on the off-diagonal.
Inserting eqs.(\ref{pM+},\ref{pM-})
it is easy to verify that:
\begin{equation}
(\bar{\nu}_{pM-} \bar{\nu}_{pM+}) M_t^{pM} \left( \begin{array}{c} \nu_{pM-} \\ \nu_{pM+}\end{array} \right) = L_{m}^{Dirac}  + L_m^{Majorana} 
\label{Diracm2}
\end{equation}
with 
\begin{equation}
M_t^{pM} = \left( \begin{array}{cc}
m_M &   m_D\\   m_D & m_M \\
\end{array} \right)
\label{massp}
\end{equation}
for $L_{m}^{Dirac}$ and $L_m^{Majorana}$ from eqs.(\ref{dirmaj},\ref{majform2}).
This mass matrix (or a similar one, typically
with the upper diagonal element set to 0)
is often presented in the 
literature with the claim that 
its mass eigenstates for m$_D$ $\ll$ m$_M$, that are
${\nu}_{pM+}$ and ${\nu}_{pM-}$  to good approximation, 
are Majorana states. The origin of
this misconception lies in the first misconception
about charge conjugation, discussed in section \ref{defins}:
It is evident e.g. from eq.(\ref{pM+}) that $\nu_{pM+}$
would be self-charge conjugate if charge
conjugation would flip chirality, but this is not the case.
The fact that both components of $\nu_{pM+}$ have SM
weak interaction might have led to a confusion 
that it conforms to the confusion theorem (section \ref{confuth}).
Summarising three bases have been
discussed: ``active-sterile'' basis of pseudo-Majorana states
$\nu_{pM+}$ and $\nu_{pM-}$, the ``particle-antiparticle'' basis
of Dirac states $\nu_D$ and $\nu^c_D$ and the ``charge-parity'' basis
of Majorana state $\nu_+$ and $\nu_-$.
All of them can be used to study the structure of the neutrino field.
But they must not be confused. 

\subsubsection{Small m$_M$ $\leq$ m$_D$: pseudo-Dirac neutrinos}
\label{dirstate}
This case is similar to the one discussed in the previous subsection \ref{pM},
but there is no relative phase between $\nu_+$ and $\nu_-$ because the
sign of the mass terms in eq.(\ref{majopl}) and eq.(\ref{majomi}) is the same:
\begin{equation}
{\nu}_{pD} = {1 \over \sqrt{2}}(\nu_+ + \nu_-) \neq {\nu}_{pD}^c.
\end{equation}
$\nu_{pD}$ the well known pseudo-Dirac state\cite{wolfenstein}, 
proposed by Pontecorvo in the very first
paper on neutrino oscillations\cite{ponte}.
\\
Upon production pseudo-Dirac neutrinos are Dirac neutrinos.
But because m$_+ \neq m_-$
the $\nu_+$  and $\nu_-$ components of a pseudo-Dirac neutrino
have different momenta and develop a relative phase
of e$^{i \beta}$ with $\beta$ = $\pi$ D $\Delta$m$^2$/(2 E) 
after propagating a distance D (with $\Delta m^2 = m_+^2 - m_-^2)$.
The phase angle is $\beta$=$\pi$ when:
\begin{equation}
D(\beta=\pi) = L = 2 {E \over \Delta m^2} = {E \over {2 m_D m_M}}.
\label{length}
\end{equation}
The pseudo-neutrino has then oscillated into a pseudo-antineutrino
\begin{equation}
{\nu}_{pD}(\beta = \pi) = {1 \over \sqrt{2}}(\nu_- + e^{i\pi} \nu_+) =  
{1 \over \sqrt{2}}(\nu_+ - \nu_-) = {{\nu}_D}^c.
\end{equation}
These oscillations violate LNC, a lepton oscillates
into an anti-lepton that can then effect a SM
weak reaction with an right-chiral amplitude suppressed
by a=(${m_- \over (E+p)})^2$. Such an anti-lepton is sterile to first
approximation but violates LNC with a typically very small reaction
amplitude a.
\\
Neutrinoless double $\beta$-decay is strongly suppressed 
for pseudo-Dirac neutrinos because here
the initial weakly produced
neutrino strictly conserves lepton number.
It then oscillates and violates LNC
with a wavelength L
(eq.(\ref{length})) that is 
very large compared to nuclear dimensions r$_N$. A very rough
estimate assumes that an emitted neutrino oscillates over the
a distance D=r$_N$ developing a small antineutrino amplitude
that then annihilates with another emitted neutrino. Approximating
r$_N$ $\approx$ 0.1/E (for energies on the order of a few MeV), it
yields a suppression factor relative to the
decay rate induced by a Majorana neutrinos
of ${1 \over 2}({r_N \over L})^4$ = ${10^{-4} \over 512}({\Delta m \over E})^8$.
Assuming ${\Delta m \over E} < 10^{-5}$ and m$_-$ = 1 eV,
this yields a completely negligible
lower limit on the pseudo-Dirac neutrino induced neutrinoless double-beta decay
half life of $^{76}$Ge (assuming a Majorana-neutrino induced
half life of 10$^{24}$ years for m$_M$=1 eV\cite{kayser})
on the order of $>$ 10$^{70}$ years.
If the physical neutrino were a pseudo-Dirac neutrino
an observation of neutrinoless double-beta decay
would be evidence for new physics beyond the
SM {\it and} beyond the mere addition of Majorana mass terms.

\subsubsection{Large m$_D$ $\simeq$ m$_M$:
the Majorana field}
\label{largm}
In essence Majorana's insight was that
the two components 
$\nu_+$ and $\nu_-$ of a Dirac field 
are dynamically independent,
i.e. they are both solutions to the Dirac
equation.
Therefore if one of the two components
would have a mass so large that its particles
are not produced in experiments and the 
other had a small mass a Dirac field
would only create Majorana particles and
thus effectively be a Majorana field.
\\
Such a scenario is realised if m$_D$ and m$_M$
are both larger than the reaction energy of a given
experiment 
so that particles of the $\nu_+$ field
have a too large mass m$_+$ = m$_D$ + m$_M$ 
(eq.(\ref{mass2})) to
appear in the laboratory and 
- at the same time - m$_M$ is equal to m$_D$
within to better than an eV or so, so that
the physical Majorana neutrino $\nu_-$ 
has a small mass m$_-$ = m$_D$ - m$_M$ $<$ eV, that
does not
violate experimental bounds on neutrino masses.
Such a scenario seems contrived, because there is no
known reason why the Majorana mass should be
on the weak scale and nearly exactly equal to the Dirac mass
but it cannot
be excluded that there is some theoretical
reason for such a fine tuning\footnote{Another possibility
would be Majorana masses that depend on charge-parity,
so that m$_{+}$ $\gg$ eV $>$ m$_{-}$.}.
\\
If a SMI-neutrino would be such a Majorana neutrino,
what would be its interactions? 
We had concluded in section \ref{confuth}, that
the historic confusion theorem no longer tells us 
that they must be the same as the ones for a Dirac neutrino
for m $\rightarrow$ 0 because this theorem rests on the
assumption PE.1 that states that a Majorana neutrino does not interact
with SM-model weak interactions but the exotic lepton
number violating Pauli-Enz interaction.
Therefore we have to calculate the transition element 
from a positron and W$^-$ state
to an
an approximately right-chiral Majorana neutrino
as in eq.(\ref{tran1r}) with
SM interactions:
\begin{eqnarray}
{\langle (e^+ W^-)_k | H_{SM}| \bar{\nu}_{+R} \rangle } =
\nonumber
\\
{1 \over \sqrt{2}}\left(\langle (e^+ W^- )_k | H_{SM}| \bar{\nu}_{DR} \rangle +
\langle (e^+ W^-)_k | H_{SM}| {\bar{\nu^c}}_{DR} \rangle\right) =
c/\sqrt{2}
\label{tran3}
\end{eqnarray}
Here
$\bar{\nu}_{+}$ was decomposed according to eq.(\ref{cmaje+}),
the value c of $\langle (e^+ W^-)_k | H_{SM}| \bar{\nu}_{DR} \rangle$
was taken from eq.(\ref{tran1r}) and the
antineutrino component 
\begin{equation}
\langle (e^+ W^-)_k | H_{SM}| \bar{\nu^c}_{DR} \rangle = 0
\end{equation}
from
eq.(\ref{intsm}). 
The transition amplitude is a factor 1/$\sqrt{2}$ smaller
than the one of eq.(\ref{tran1r}).
\begin{itemize}
\item
{\bf Fact}
\\
{\it A Majorana neutrino with SM-interactions would be produced with a cross
section 1/2 times smaller than expected for SMI-Dirac
neutrinos. Therefore the hypothesis that the physical neutrino is
a Majorana neutrino 
is in quantitative contradiction to
experimental evidence.}
\end{itemize} 
The reduction would apply in an analogous
manner also to neutral current reactions.
This excludes that the physical neutrino is in the state $\nu_-$
(or $\nu_+$).
\\
This effective ``reduction'' in weak transition amplitude due to a mass
occurs also in other familiar circumstances. In an analogous manner, a non-relativistic
neutrino (or electron) produced at rest has half the weak transition amplitude
squared
compared to one produced at ultra-relativistic 
energies because it has a sterile, right-chiral
component with the same amplitude as the active left-chiral amplitude.
A Dirac antineutrino at rest and a massless Majorana neutrino emitted
in a $\beta$-decay together with an electron, both have sterile left-chiral
and an active right-chiral amplitude of value ${1\over{\sqrt{2}}}$.
\\
Summarising, Majorana neutrinos with SM interactions need a fine-tuned
combination of Dirac and Majorana mass terms and 
have a phenomenology
that quantitatively differs from the one of the observed neutrino
in having an effective weak transition amplitude $\sqrt{2}$ smaller
than the observed one.

\subsection{Summary of the restriction mechanisms, comparison with experiment}
\label{rsum}
Table \ref{t1} summarises the diverse mechanisms that restrict
or enlarge the set of states
that a SMI-neutrino can be in, thus creating various
LNC violating neutrino types. Pseudo-Dirac neutrinos
exclude no states but {\it include} states that the SMI-neutrino
can normally not be in, namely the right(left)-chiral (anti)neutrino. 
\\
Experimental evidence about physical SMI-neutrinos was found to agree
only with Dirac or pseudo-Dirac neutrinos: pseudo-Majorana neutrinos have a qualitatively
and Majorana neutrinos a quantitatively different phenomenology.
Based on this insight I formulate the following
\begin{itemize}
\item
{\bf Conjecture}
\\
{\it The physical neutrino is probably a
pseudo-Dirac neutrino because it
seems very likely that small Majorana
masses are induced  by some see-saw mechanism\cite{weinberg}
and a larger Dirac mass is then needed to predict
the observed phenomenology for a SMI-neutrino.}
\end{itemize}
\begin{table}
\begin{tabular}{llll}
{\bf Field:} & {\bf Excl.}: & {\bf Incl.}: & {\bf Reason for exclusion or inclusion}: \\ 
{ No Maj. mass}: & - & - & - \\ 
{ Interaction-ind. Maj.}:  & $\nu_-$ & - & $\nu_-$ not part of weak-interaction theory \\ 
{ Mass-ind. Majorana}: & $\nu_-$ & - & $\langle\nu_-|\nu_p\rangle = 0$ for E($\nu$)$\ll$ 100 GeV \\ 
{ Pseudo Majorana}: & $\nu_D,\nu_M,\nu_L,\nu_R$ & - &  Phase diff. $\gamma_5$ btw. $\nu_+$ and $\nu_-$ \\ 
{ Pseudo-Dirac}: & - & $\nu_{DR},\bar{\nu}_{DL}$ & Different momentum of $\nu_+$ and $\nu_-$
\end{tabular}
\caption{\it 
Weak-interaction induced Majorana (second row, section \ref{winter}), mass-induced Majorana (third row,
section \ref{largm}),
pseudo-Majorana (fourth row, section \ref{pM}) and pseudo-Dirac neutrino (fifth row, section \ref{dirstate})
have different mechanisms 
that exclude or include the states from the set of states the original SMI-neutrino (first row) can be in.
``Excl.'' stands for excluded states, ``Incl.'' for included states relative to the 
original Majorana-massless SMI-neutrino. $\nu_p$ is the state of the physical neutrino.}
\label{t1}
\end{table}
\section{Off-axis $\mathcal{CPT}$-violating masses}
\label{offaxis}
In the charge-parity basis (eq.(\ref{vecN})), 
the mass matrix remained diagonal even with
Majorana masses (eq.(\ref{mass2})). It is interesting to ask what happens
when we add
off-diagonal terms m$_O$ so that:
\begin{equation}
M = \left( \begin{array}{cc}
m_+ &   m_O\\   m_O &  m_- \\
\end{array} \right)
\label{mass3}
\end{equation}
The mass eigenstates of this matrix are no
longer the Majorana states $\nu_{+,-}$. 
In order to find the eigenstates $\nu^{'}=U \nu_D$ 
the matrix M must be diagonalised with a unitary
matrix $U$ = {\rm $\left( \begin{array}{cc}
\rm{cos}(\theta) &   \rm{sin}(\theta)\\   \rm{-sin}(\theta) & \rm{cos}(\theta) \\
\end{array} \right)$}
such that M$^{'}$ = $U M U^{-1}$ is a diagonal
matrix:
\begin{equation}
M^{'} \nu^{'} = \left( \begin{array}{cc}
m_+^{'} &   0\\   0 & m_-^{'} \\
\end{array} \right) 
\left( \begin{array}{c} \rm{cos}(\theta) \nu_+ + \rm{sin}(\theta) \nu_- \\ \rm{cos}(\theta) \nu_+ - \rm{sin}(\theta) \nu_-  \end{array} \right).
\label{mass3a}
\end{equation}
One finds
the following expression for the mixing angle
\begin{equation}
{\rm \theta} = {{{\rm arctan}\left({{2 m_o} \over {m_+ - m_-}}\right)}\over 2} =
{{{\rm arctan}\left({{m_o} \over {m_M}}\right)}\over 2}.
\label{angle}
\end{equation}
and for the eigenmasses
\begin{equation}
m^{'}_{+,-} = 1/2 \left( m_+ + m_- \pm \sqrt{(m_+ + m_-)^2 - 4 (m_+ m_- - m_O^2)}\right).
\label{masses}
\end{equation}
m$^{'}_-$ is positive if m$_+$m$_-$ $\geq$ m$_O^2$. 
Up to this section I considered the case m$_O$ = 0 that yields $\theta$=0. 
If m$_O$ $\neq$ 0
there is maximal mixing i.e. $\theta$ = 45$^o$ if
m$_+$ $\rightarrow$ m$_-$ i.e. m$_M$ $\rightarrow$ 0.
According to eqs.(\ref{dirac1},\ref{cdirac})
in this case
$\nu^{'}$ = $\left( \begin{array}{c} \nu_D \\ \nu_D^c \end{array} \right)$,
i.e. the mass eigenstates are the Dirac particle and anti-particle state.
This means that particle and antiparticle have different
masses and thereby the mass violates $\mathcal{CPT}$ invariance.
For intermediate angles of $\theta$ between 0 and $\pi/4$
the mass-eigenstates are mixtures between Dirac and pseudo
states (pseudo-Dirac if m$_-^{'}$ $\geq$ 0 and pseudo-Majorana if  m$_-^{'}$ $<$ 0
\footnote{Only the pseudo-Dirac case is studied below, because
pseudo-Majorana states do not have the observed phenomenology (section \ref{pM}),
i.e. we assume that m$_+$m$_-$ $\geq$ m$_O^2$.}), i.e. $\mathcal{CPT}$ is still violated.
Therefore one might exclude the existence of
non-vanishing masses off-diagonal in the
charge-parity basis
because they are incompatible with basic assumptions on which
quantum field theory is built\cite{okun}.
However, the possibility of $\mathcal{CPT}$ violation in the
neutrino sector has been widely discussed\cite{barenboim,giunti,cpt}, and it
is interesting to characterise the precise meaning of
the mixing angle $\theta$.
With finite $\theta$
SM interactions could produce a neutrino together with a positron
as a mixture of Dirac/pseudo Dirac neutrino:
\begin{equation}
{\nu}_{D-pD} = {\rm cos}(\pi/4-\theta)\nu_D + {\rm sin}(\pi/4-\theta)\nu_{pD}
\label{neus}
\end{equation}
The pseudo-Dirac amplitude following from this expression
is A$_{pD} = \rm{sin}^2(\pi/4-\theta)$. Thereby $\theta$
determines the pseudo-Dirac amplitude with which a neutrino with
Majorana and off-axis mass terms oscillates into a
sterile state. This mechanism is therefore an interesting -
if exotic - alternative to new sterile neutrino generations
that mix with active generations to interpret the tentative 
experimental evidence for sterile neutrinos\cite{sterile}.
\\
Such a scenario is predictive. If it were
possible to determine m$_+$, m$_-$ and m$_O$ by
measuring the mass of neutrino and antineutrino
(eq.(\ref{masses})) and the mixing angle
of sterile neutrinos (eq.(\ref{angle})) then from
m$_M$ = (m$_+$ - m$_-$)/2 and m$_D$ = (m$_+$ + m$_-$)/2, the 
wavelength of the neutrino - sterile neutrino oscillation
could be predicted via eq.(\ref{length}).
\section{Summary of the novel theoretical results}
\label{sum}
If the standard model describes the weak interaction
correctly to good approximation, the only way to
make the neutrino 
a Majorana neutrino is by way of mass terms
(section \ref{winter}).
If there is a small Majorana mass term that
is larger than the Dirac mass term, the neutrino field
consists of a neutrino and antineutrino component
with opposite chirality (pseudo-Majorana neutrino, section \ref{pM}).
If there
is a large Dirac mass term and a large Majorana mass
term of nearly equal value the neutrino is a Majorana
neutrino. Such a neutrino has a SM weak interaction 
transition amplitude 
that is half as large as the experimentally
observed one (section \ref{largm}).
The neutrino is a pseudo-Dirac neutrino
not only if a small Dirac mass 
is {\it much} larger than a finite Majorana mass,
but also if it is just larger or of equal value.
(section \ref{dirstate}). 
Mass terms that couple field states with opposite charge parity violate
$\mathcal{CPT}$ invariance (with different neutrino and antineutrino
masses) (section \ref{offaxis}).

\section{Conclusions about the physical neutrino}
\label{concl}
Physical neutrinos have a definite helicity to good
approximation and can therefore not be pseudo-Majorana neutrinos.
The precise quantitative agreement of the neutrino's phenomenology
with the properties expected in the standard model\cite{giunti} 
makes it impossible that {\it any} of the three
neutrino field is a Majorana field for
which it would be expected that the squared weak 
transition amplitude squared
is half as large as expected in the SM(section \ref{largm}).
Under the 
assumption that the physical neutrino
is a SMI-neutrino to good approximation (which is universally
made in high-energy physics)
it is therefore certain that:
\begin{itemize}
\item
 m$_D$ $>$ m$_M$,
\item
 the
physical neutrino is either a Dirac or pseudo-Dirac particle,
\item
 neutrinoless double-$\beta$ decay with a non negligible
rate cannot be due only to a Majorana mass term.
\end{itemize}
Moreover because of the observational fact that neutrinos oscillate among 3 generations\cite{giunti}, 
at least two neutrino generations must have a finite Dirac mass.
\\
It is very likely that neutrinos are pseudo-Dirac particles  
because there is no plausible reason why the 
Majorana mass of the neutrino should
vanish exactly.
This is good news,
because with $\mathcal{CPT}$ invariance they oscillate from neutrino to antineutrino
either with maximal amplitude, or, if $\mathcal{CPT}$ invariance is broken (which would
be even more interesting), with a smaller
amplitude. 
Pseudo-Dirac neutrinos offer excellent prospects
for a definite discovery of even extremely small Majorana mass terms
because even
extremely large oscillation lengths in principle can be probed with current
technology\cite{posci}. 
Non-negligible
neutrinoless double-beta decay assumes a role exclusively as a channel
to search for new physics {\it beyond} the mere addition of
Majorana mass terms (e.g.\cite{gu}). 

\section*{Acknowledgments}
I thank Palash Pal and Carlo Giunti for recent
inspiring and helpful discussions on the subject
of Majorana neutrinos and Silvia Pezzoni
for convincing me that 18 years ago - contrary
to my then firm conviction -
I did not understand
the nature of Majorana neutrinos at all.

\section{Appendices}
\label{app}
\subsection{The Majorana representation of $\gamma$ matrices}
\label{rep}
I choose the following basis of $\gamma$ matrices:
\begin{eqnarray}
\gamma^0 = \left( \begin{array}{cc}
0 &   \sigma^2\\   \sigma^2 & 0 \\
\end{array} \right)
\nonumber
\\
\gamma^1 = \left( \begin{array}{cc}
0 &   \sigma^1\\   \sigma^1 & 0 \\
\end{array} \right)
\nonumber
\\
\gamma^2 = \left( \begin{array}{cc}
\bf{1} \ & 0  \\  0 & - \bf{1} \ \\
\end{array} \right)
\nonumber
\\
\gamma^3 = \left( \begin{array}{cc}
0 &   \sigma^3 \\   \sigma^3 & 0 \\
\end{array} \right)
\end{eqnarray}
Itzykson \& Zuber\cite{itz} choose a slightly different
form in their appendix, with which the definition
of charge conjugation (my eq. (\ref{def2})) is:
\begin{equation}
{{\nu}}(x,t)^c = -i \ {{\nu}}(x,t)^{\dagger T}  \ \ \ ({\rm alternative \ Majorana \ representation})
\label{defi}
\end{equation}

\subsection{Can a weakly interacting neutrino be self-charge ($\mathcal{C}$) conjugate?}
\label{app1}

It has been argued\cite{kayser82,kayser} that ``dressed'', i.e. weakly interacting
neutrinos cannot be self-charge conjugate
i.e. cannot fulfil eq.(\ref{def}).
This ``argument C'' runs like this\cite{kayser}:
\\
{\it ``A predominantly left-chiral Majorana neutrino can virtually decay into a
e$^-$ W$^+$ pair but not a e$^+$ W$^-$ pair due to the
weak interaction's charge-parity
violation. e$^-$ W$^+$ is not an eigenstate of $\mathcal{C}$
and therefore the Majorana neutrino is neither,
because it has maximally $\mathcal{C}$-violating
interactions.''}
\\
The final decay state e$^-$ W$^+$ is indeed not an eigenstate of $\mathcal{C}$ and
therefore has no definite charge-parity. But from this fact
one can only infer the charge-parity of the initial Majorana neutrino
if the interaction were charge-parity conserving.
\\
Let us consider a different, charge-parity conserving reaction where 
a similar inference can be drawn correctly:
the decay $\pi_0$ $\rightarrow$ 2 $\gamma$.
The charge parity of the final state is +1,
because the photon has a charge parity of -1 ( a fact derived from
the structure of an electromagnetic quantum field\cite{bd}).
Because the decay proceeds via the electromagnetic 
interaction which conserved charge-parity, one can
infer that the $\pi_0$ has a charge parity of +1.
\\
On the contrary, the decay $\nu_M$ $\rightarrow$ e$^+$ W$^-$ proceeds
via the weak interaction which does not conserve
charge parity so no conclusion about the charge
parity of the neutrino can be drawn from the charge
parity of the final state, i.e. the above ``argument C''
is wrong.
\\
The inference of the charge parity of a field from
the one of another in decay processes needs always
to be rooted in the determination of the charge
parity of some field from the structure of the 
quantum field (as was done for the photon\cite{bd} in the
above example). This rooting is done in eqs.(\ref{majc1}) to
(\ref{dirac1}) for the case of the Majorana fields that
make up a Dirac quantum field.
The fact that for an only weakly interacting
field, like the neutrino, one can draw no conclusions about
the charge parity of decay products, does
not invalidate this rooting analysis.
\\
Another angle on this problem is the following.
Is a weakly interacting Dirac neutrino in a self-parity ($\mathcal{P}$) conjugate
state? For a massless neutrino the answer is ``no,
because the weak interaction violates parity''. 
The parity conjugate of a left-handed neutrino
is a right-handed neutrino, i.e. it is not self-parity conjugate.
We saw in section \ref{winter} that a massless SMI-neutrino 
is also not self-charge conjugate.
\\
But masses can change these conclusions.
In the limit m$_D$ $\rightarrow$ E, where E is the
reaction energy, Dirac mass terms can
``make'' the neutrino
self-parity conjugate (see section \ref{pcon} for 
a detailed discussion), even though it
is produced in a reaction that still violates $\mathcal{P}$-parity
(as indicated by the properties of other particles
taking part in the reaction).
\\ 
In a similar manner Majorana
and Dirac masses make the neutrino self-charge conjugate (as
outlined in section \ref{largm})
even if it was produced in a reaction that
violates charge parity. 
Therefore one cannot conclude from the 
noninvariance of the reaction-Hamiltonian with respect
to $\mathcal{C}$ or $\mathcal{P}$ that some fermion fields taking 
part in the reaction are not made self-charge or
self-parity invariant by special mass terms.
\\
Nor can one conclude from the invariance of a
reaction-Hamiltonian to $\mathcal{C}$ that the fermion fields
of the reaction products are 
self-charge conjugate, consider e.g. the electromagnetic 
reaction $\gamma$ $\rightarrow$ e$^+$ + e$^-$.
The concepts ``$\mathcal{C}$-invariance of reaction-Hamiltonian''
and ``$\mathcal{C}$-invariance of a field taking part in the reaction''
are logically independent concepts.
\\
{\it Summarising, weakly interacting, neutral fields
with spin 1/2 can in principle be self-charge ($\mathcal{C}$)
conjugate.}

\subsection{How a Dirac mass term can make a neutrino field self-parity conjugate}
\label{pcon}
A massless neutrino that was produced by H$_{SM}$
\begin{equation}
{\nu_{DL}} = {\int\limits_{k}}^{(+)} b_k^{\dagger} v_L(x) + {\int\limits_{k}}^{(-)} d_k u_R(x)
\end{equation}
is not self-parity conjugate
because the parity conjugation operation $\mathcal{P}$ flips chirality, i.e. $\gamma_0 v_L(-x) = - v_R(x)$,
$\gamma_0 u_L(-x) =  u_R(x)$
but does not change $b$,$d^\dagger$ and only changes the sign of $d$,$b^\dagger$\cite{bd}, so
that:
\begin{equation}
\mathcal{P}{\nu_{DL}}\mathcal{P}^{-1} = {\int\limits_{k}}^{(+)} b_k^{\dagger} v_R(x) + {\int\limits_{k}}^{(-)} d_k u_L(x) \neq {\nu_{DL}}
\end{equation}
However, it is well known (e.g.\cite{palash}), that
if the neutrino has a Dirac mass m$_D$ $\rightarrow$ E,
i.e. the decay is non-relativistic (NR), 
the two chiralities are produced and annihilated with
equal amplitude i.e.
\begin{equation}
{\nu_{NR-D}} = {\int\limits_{k}}^{(+)} b_k^{\dagger} (v_L(x)+v_R(x)) + {\int\limits_{k}}^{(-)} d_k  (u_L + u_R(x))
\end{equation}
and clearly:
\begin{equation}
\mathcal{P} \nu_{NR-D} \mathcal{P}^{-1} = \nu_{NR-D}
\end{equation}
The field was made self-parity conjugate by the Dirac mass
in the limit m$_D \rightarrow$ E,
because it excluded states with a definite chirality of 1 or -1.

\subsection{Weak reactions with H$_{PE}$ vs. H$_{SM}$}
\label{lncd}
If H$_{PE}$ would be the weak-interaction Hamiltonian
weak reactions would violate LNC.
We can illustrate a LNC reaction that produces neutrinos:
\begin{equation}
e^+ W^- \rightarrow \bar{\nu}_D 
\label{Dreac}
\end{equation}
and
\begin{equation}
e^+ W^- \rightarrow \bar{\nu}_+.
\label{Mreac}
\end{equation}
The left-hand side of both equations has a lepton number
of -1 (of the positron).
The right-hand side of eq.(\ref{Dreac}) (a Dirac neutrino) also has a lepton number of $\ell$ = -1 because
\begin{equation}
\mathcal{L} {\bar{\nu}_D} = - {\bar{\nu}_D} 
\label{lnumber1}
\end{equation}
where  $\mathcal{L}$ is the lepton-number operator.
Therefore lepton number is conserved in eq.(\ref{Dreac}).
However, the right-hand side of eq.(\ref{Mreac})
(a Majorana neutrino) has no well defined lepton number
because from eq.(\ref{lnumber1}),
\begin{equation}
\mathcal{L} \nu_D = \nu_D 
\label{lnumber2}
\end{equation}
and  eq.({\ref{cmaje+}})
it follows that the Majorana
neutrino is no eigenstate of the lepton-number operator $\mathcal{L}$:
\begin{equation}
\mathcal{L} \nu_+ = {1 \over \sqrt{2}} (\nu - \nu^c) \neq \ell \nu_+.
\end{equation}
The lepton number $\ell$ of $\nu_+$ is undefined, 
rather then being close to -1, which would
be necessary if the SM described the interaction
to good approximation. 
LNC in eq.(\ref{Mreac}).
Therefore the reaction eq.(\ref{Mreac}) does not occur for
a massless (or Dirac-massive) neutrino with H$_{SM}$.

\end{document}